\documentstyle[12pt]{article}
\newcommand{\vj}{{\bf{j}}} %-vector j
\newcommand{\vx}{{\bf{x}}} %-vector x
\newcommand{\bfnabla}{{\mbox{\boldmath $\nabla$}}} %-vector nabla
\newcommand{\bfdd}{{\mbox{\boldmath $\partial$}}} %-vector partial

\newcommand{\Sum}[1]{\displaystyle{\sum_{#1} \kern-1.20em \int}} 

\newcommand{\io}{[\hspace{-1pt}[}
\newcommand{\ic}{]\hspace{-1pt}]}
\newcommand{\minlam}{-\hspace{-4pt}\lambda}

\renewcommand{\Re}{{\rm Re}}

\def\Arch{{\rm Arcosh}}

\date{}

\begin {document}

\begin{titlepage}
\centerline{\LARGE
The Casimir-Aharonov-Bohm effect?}
\vspace{1cm}
\begin{center}
{\large Yu.A.~Sitenko and A.Yu.~Babansky}\\
\mbox{} \\
{\it Bogolyubov Institute for Theoretical Physics,}\\
{\it National Academy of Sciences of Ukraine,}\\
{\it  252143 Kyiv, Ukraine}
\end{center}

\vskip 1 cm
\begin{abstract}
The combined effect of the magnetic field background
in the form of a singular vortex and the Dirichlet boundary condition
at the location of the vortex on the vacuum of quantized scalar field
is studied. We find the induced vacuum energy density and current to be
periodic functions of the vortex flux and holomorphic functions of the
space dimension.
\end{abstract}

\medskip


\vfill

\end{titlepage}

\setcounter{page}{1}

The emergence of calculable and detectable vacuum energy in quantum 
field theory under the influence of boundary conditions was predicted first by
Casimir \cite{Cas}. Let $X$  be the base space manifold and $Y$ be a submanifold
of dimension less than that of $X$. Then the matter field in $X$ is
quantized  under a certain boundary condition imposed at $Y$ and the
vacuum polarization effects are studied. Usually $Y$ is chosen to be
noncompact disconnected (e.g.
two parallel infinite plates, as generically in Ref. \cite{Cas}) or closed
compact (e.g. box or sphere), see Refs. \cite{Mos} and \cite{Eli}. 
However, as it seems to us, one possibility has been overlooked, namely
that $Y$ is noncompact connected. Therefore we propose to consider such a
possibility, moreover to choose $Y$ as a manifold that confines
a singular magnetic vortex in itself: this will allow us to find one
more manifestation of the Aharonov-Bohm effect \cite{AhaR} in vacuum
polarization \cite{Sit90}.

To be more precise, we start from the operator of the secondly quantized
complex scalar field in an external  static background
\begin{equation}
\Psi(\vx,t)=\Sum{\lambda}\;\frac{1}{\sqrt{2E_\lambda}}
\left[e^{-iE_\lambda t}\,\left<\vx|\lambda\right>\,a_\lambda+
e^{iE_\lambda t}\,\left<\vx|\minlam\right>\,{b_\lambda}^{+}\right],
\label{form1}
\end{equation}
where ${a_\lambda}^{+}$ and $a_\lambda$ (${b_\lambda}^{+}$ and $b_\lambda$)
are the scalar particle (antiparticle) creation and annihilation
operators satisfying the commutation relation
\begin{equation}
\Big[a_\lambda,{a_{\lambda'}}^{+}\Big]_{-}
=\Big[b_\lambda,{b_{\lambda'}}^{+}\Big]_{-}
=\left<\lambda|\lambda'\right>,
\label{form2}
\end{equation}
$\lambda$ is the set of parameters (quantum numbers) specifying the state,
$E_\lambda=E_{-\lambda}>0$ is the energy of the state; symbol  $\Sum{\lambda}$ implies
the summation over the discrete and the integration
(with a certain measure) over the continuous values of $\lambda$.
The wave function $\left<\vx|\lambda\right>$ satisfies the stationary
Klein-Gordon equation
\begin{equation}
\left(-\bfnabla^2 + m^2\right)\,\left<\vx|\lambda\right>={E_\lambda}^2\,\left<\vx|\lambda\right>,
\label {form3}
\end{equation}
($\bfnabla$ is the covariant derivative in an external
static background) and the Dirichlet boundary condition
\begin{equation}
\left<\vx|\lambda\right>\Big|_{\,\vx\,\in\,Y} =0.
\label{form4}
\end{equation}
We take Euclidean d-dimensional space as a base space $X$ and define $Y$ 
as follows
\begin{equation}
Y\;:\;\; x^1=x^2=0,
\label{form5}
\end{equation}
i.e. the point in the case of $d=2$, the line in the case of $d=3$ and the
$(d-2)$-dimensional hypersurface in the case of $d>3$.
We take classical static magnetic field as an external background,
thus the covariant derivative is defined as
\begin{equation}
\bfnabla = {\bfdd} - i{\bf{V}}(\vx),
\label {form6}
\end{equation}
where ${\bf{V}}(\vx)$ is the vector potential of the magnetic field;
note that in the  $d$-dimensional space the magnetic field strength
is an antisymmetric tensor of the rank $d-2$
\begin{equation}
B^{\textstyle{\nu_1\cdot\cdot\cdot\nu_{d-2}}}(\vx)=
\left[\partial_{\textstyle{\mu_1}}V_{\textstyle{\mu_2}}(\vx)\right]
\epsilon^{\textstyle{\mu_1 \mu_2 \nu_1 \cdot\cdot\cdot \nu_{d-2}}},
\label {form7}
\end{equation}
$\epsilon^{\textstyle{\mu_1\cdot\cdot\cdot\mu_d}}$ is the totally antisymmetric 
tensor, $\epsilon^{1 2 \cdot\cdot\cdot d}=1$.
The magnetic field configuration is chosen to be that of a singular
vortex placed at $Y$:
\begin{equation}
V_1(\vx)=-\Phi \frac{x^2}{(x^1)^2+(x^2)^2}, \,\,\,
V_2(\vx)=\Phi \frac{x^1}{(x^1)^2+(x^2)^2}, \,\,\,
V_\nu(\vx)=0, \,\,\,
\nu=\overline{3,d}\,\, ,
\label{form8}
\end{equation}
\begin{equation}
B^{3 \cdot \cdot \cdot d}(\vx)=2\pi\, \Phi\, \delta(x^1)\, \delta(x^2),
\label{form9}
\end{equation}
$\Phi$ is the total flux (in the units of  $2\pi$) of the vortex.
In this Letter we shall find how the vacuum of quantized scalar field is
polarized under the boundary condition (\ref{form4}) and (\ref{form5})
in the magnetic field background (\ref{form6}) and (\ref{form8}).
  
The crucial role is played by the zeta function of the operator
$(-\bfnabla^2 + m^2)$
\begin{equation}
\zeta(s)=\int d^d x\;\zeta_{\vx}(s),
\label{form10}
\end{equation}
where the zeta function density can be presented in the form
\begin{equation}
\zeta_{\vx}(s)=\lim_{\vx'\to\vx}
\left<\vx\left|(-\bfnabla^2 + m^2)^{-s}\right|\vx'\right>.
\label{form11}
\end{equation}
In the free field case, i.e. in the absence of any boundary
condition and any background field, one has
\begin{equation}
\zeta_{\vx}^{(0)}(s)=\lim_{\vx'\to\vx}
\left<\vx\left|(-\bfdd^2 + m^2)^{-s}\right|\vx'\right>=
(2\pi)^{-d}\, \int d^d p\, (p^2+m^2)^{-s}.
\label{form12}
\end{equation}
The integral in Eq.(\ref{form12}) is convergent
at $s>\frac{d}{2}$ when it can be easily evaluated:
\begin{equation}
\zeta_{\vx}^{(0)}(s)
= \frac{m^{{d-2s}}}{(4\pi)^{\frac{d}{2}}}\;
\frac{\Gamma \left(s-\frac{d}{2}\right)}{\Gamma(s)},
\label{form13}
\end{equation}  
where $\Gamma(z)$ is the Euler gamma function.
We have calculated the zeta function density (\ref{form11}) under the conditions
specified in Eqs.(\ref{form4})-(\ref{form6}) and (\ref{form8})
(details will be published elsewhere):
\begin{equation}
\zeta_{\vx}(s)=\frac{4\sin(F\pi)}{(4\pi)^{{\frac{d}{2}+1}}}\,\frac{m^{{d-2s}}}{\Gamma(s)}
\int_{0}^{\infty} du\, e^{-{u}}\, \left[ K_F(u) + K_{1-F}(u) \right]\,
\gamma \left( s-\frac{d}{2},\; \frac{m^2 r^2}{2u} \right),
\label{form14}
\end{equation}
where
\begin{equation}
\gamma(z, w) = \int_{0}^{w} du\,u^{z-1}\,e^{-u}
\label{form15}
\end{equation}
is the incomplete gamma function and $K_\omega(z)$ is the
McDonald function of the order $\omega$, $r = \sqrt{(x^1)^2 + (x^2)^2}$
and
\begin{equation}
F=\Phi - \io \Phi \ic, \qquad 0 \leq F < 1,
\label{form16}
\end{equation}
$\io w \ic$ is the integer part of the quantity $w$ (i.e. the
integer which is less than or equal to $w$); note that 
$\zeta_{\vx}(s)$ (\ref{form14}) is a periodic
function of the vortex flux $\Phi$, since it depends only on $F$
(being symmetric under $F\to 1-F$).

Although $\zeta_{\vx}(s)$, as well as $\zeta_{\vx}^{(0)}(s)$,
has been calculated at $s>\frac{d}{2}$, both results (\ref{form13}) and
(\ref{form14}) can be extended analytically to the whole complex s-plane;
for $\zeta_{\vx}^{(0)}(s)$ the analytic continuation is given obviously by the
gamma function, while for $\zeta_{\vx}(s)$ the analytic continuation is
obtained with the help of the incomplete gamma function (\ref{form15}),
i.e. using repeatedly the recurrence relation
\begin{equation}
\gamma(z, w) =
\frac{1}{z} \left[ \gamma(z + 1, w) + w^z e^{-w} \right].
\label{form17}
\end{equation}
Moreover, both products $\zeta_{\vx}(s)\Gamma(s)$ and
$\zeta_{\vx}^{(0)}(s)\Gamma(s)$ have the same singularity structure
comprising of simple poles on the real axis at
$s = \frac{d}{2}+1-N\,(N = 1,2...)$. Therefore, the difference
\begin{equation}
\zeta^{ren}_\vx(s) = \zeta_\vx(s)-\zeta^{(0)}_\vx(s)
\label{form18}
\end{equation}
and even the product $\zeta^{ren}_\vx(s)\Gamma(s)$ appear to be holomorphic
on the whole complex s-plane.

The renormalized zeta function density (\ref{form18})
can be presented in the form 
\begin{eqnarray}
\zeta^{ren}_\vx(s)&=&-\frac{16\sin(F\pi)}{(4\pi)^{{\frac{d}{2}+1}}\Gamma(s)}
\left( \frac{r}{m} \right)^{{s-\frac{d}{2}}}\times
\nonumber\\
&\times& \int_1^\infty \frac{d v}{\sqrt{v^2-1}}\,\cosh \left[ (2F-1)\Arch v \right]\,
v^{{s-\frac{d}{2}-1}}\,K_{{s-\frac{d}{2}}}(2mr v).\;
\label{form19}
\end{eqnarray}
Unlike $\zeta_\vx(s)$ (\ref{form14}) and  $\zeta^{(0)}_\vx(s)$ (\ref{form13}),
$\zeta^{ren}_\vx(s)$ is decreasing exponentially
at large distances from the vortex
\begin{equation}
\zeta^{ren}_\vx(s) =-\frac{\sin(F\pi)}{(4\pi)^{{\frac{d}{2}}}\Gamma(s)}e^{-{2mr}}
m^{{\frac{d}{2}-s-1}}r^{{s-\frac{d}{2}-1}}
\left\{
1+ O \left[ (mr)^{-1} \right]
\right\}, \quad mr \gg 1. 
\label{form20}
\end{equation}
At small distances $\zeta^{ren}_\vx(s)$ is characterized by the power
behaviour
\begin{eqnarray}
\zeta^{ren}_\vx(s)=-\frac{4\sin(F\pi)}{(4\pi)^{\frac{d+1}{2}}\Gamma(s)}
\frac{\Gamma\left(\frac{d}{2}-s+F\right)\Gamma\left(\frac{d}{2}-s+1-F\right)}
{(d-2s)\Gamma\left(\frac{d+1}{2}-s\right)}\, r^{2s-d} 
\left\{ 1+O\left[(mr)^2\right]\right\},
\nonumber\\mr\ll 1.\quad
\label{form21}
\end{eqnarray}

Integrating over the plane which is orthogonal to the
vortex, we get
\begin{equation}
\int_{-\infty}^{\infty}dx^1 \int_{-\infty}^{\infty} dx^2 \; \zeta^{ren}_\vx(s)
= - \frac{m^{d-2s-2}}{2(4\pi)^{\frac{d}{2}-1}}
\frac{\Gamma(s-\frac{d}{2}+1)}{\Gamma(s)}\,F(1-F);
\label{form22}
\end{equation}
the integral in Eq.(\ref{form22}) is evaluated at $\Re\;s>\frac{d}{2}-1$,
however the result can be extended analytically to the whole complex s-plane.

The renormalized vacuum energy density is defined as
(see Eqs.(\ref{form11}),(\ref{form12}) and (\ref{form18}))
\begin{equation}
\varepsilon^{ren}(\vx)= \zeta^{ren}_{\vx}\left(-\frac{1}{2}\right),
\label{form23}
\end{equation}
thus we get
\begin{eqnarray}
\varepsilon^{ren}(\vx) &=&\frac{16\sin(F\pi)}{(4\pi)^{\frac{d+3}{2}}}
\left( \frac{m}{r} \right)^{\frac{d+1}{2}} \times
\nonumber\\
&\times&\int_1^\infty \frac{d v}{\sqrt{v^2-1}} \cosh \left[ (2F-1)\Arch v \right]
v^{-\frac{d+3}{2}}K_{\frac{d+1}{2}}(2mrv).
\label{form24}
\end{eqnarray}
The asymptotics of $\varepsilon^{ren}(\vx)$ at large and small distances
are immediately obtained from Eqs.(\ref{form20}) and (\ref{form21})
at $s=-\frac{1}{2}$. Concerning the dependence on the fractional
part of the vortex flux $F$, the renormalized vacuum energy
density is a positive function, which is convex upwards, symmetric under
$F\to 1-F$, and has maximum at $F=\frac{1}{2}$. We get 
\begin{eqnarray}
&~&\left. \varepsilon^{ren}(\vx) \right|_{F=\frac{1}{2}}=
\frac{2 m^{2N+1}} {(4\pi)^{N+\frac{1}{2}}}
\frac{(-1)^{N}}{\Gamma\left(N+\frac{3}{2}\right)}
\Bigg\{
-\frac{\pi}{2}+ \pi mr \Big[ K_0(2mr)L_{-1}(2mr)+
\nonumber\\
&~&+K_1(2mr)L_0(2mr) \Big]+\frac{1}{\sqrt{\pi}}
\sum_{l=0}^{N}(-1)^{l} \Gamma\left(l+\frac{1}{2}\right) (mr)^{-l} K_{l+1}(2mr)
\Bigg\},\;d=2N,\qquad\label{form25}
\end{eqnarray}
\begin{eqnarray}
&&\left. \varepsilon^{ren}(\vx) \right|_{F=\frac{1}{2}}=
\frac{2 m^{2(N+1)}} {(4\pi)^{N+1}} 
\frac{(-1)^{N}}{\Gamma\left(N+2\right)}
\Bigg\{
-\mbox{E}_1(2mr)+
\nonumber\\
&&+e^{-2mr}
\sum_{l=0}^{N}(-1)^{l} \Gamma\left(l+1\right)
\sum_{n=0}^{l+1}
\frac{\Gamma(l+n+2) (mr)^{-l-n-1}} {2^{2n+1}\Gamma(n+1)\Gamma(l-n+2)} 
\Bigg\},\; d=2N+1,\qquad
\label{form26}
\end{eqnarray}
where $L_{\omega}(z) $ is the modified Struve function of the order $\omega$
and
\begin{equation}
\mbox{E}_1(w) =\int_w^{\infty}\frac{du}{u}\,e^{-u}
\label{form27}
\end{equation}
is the integral exponential function (see e.g. \cite{Abra}).

Let us emphasize the following remarkable fact: since the first argument of
the incomplete gamma function in Eq.(\ref{form14}) is $s-\frac{d}{2}$
and the analytic continuation in the variable $s$ is possible, then the
analytic continuation in the variable $d$ is also possible. So,
Eq.(\ref{form24}) can be continued analytically in $d$, and the renormalized
vacuum energy density appears to be holomorphic
on the complex $d$-plane. Therefore, Eq.(\ref{form24}) can be
regarded as a result obtainable in the framework of the dimensional
regularization procedure: the difference between vacuum energy 
densities in the presence and the absence of a vortex being calculated
at $\Re\;d<-1$ and then continued analytically to the whole complex $d$-plane.

What else, in addition to the energy density, is induced in the vacuum?
Contrary to the case of the fermionic vacuum (see Refs.[5,7,8]),
the charge and the angular momentum cannot be
induced in the bosonic vacuum, so there remains the current
\begin{equation}
\vj(\vx)=\lim_{\vx'\to\vx}
\left<\vx\left|(-i\bfnabla)(-\bfnabla^2 + m^2)^{-\frac{1}{2}}\right|\vx'\right>.
\label{form28}
\end{equation}
Under the condition (\ref{form4})-(\ref{form6}) and (\ref{form8})
only one (angular) component is nonvanishing:
\begin{eqnarray}
&&j_{\varphi}(\vx)\equiv r^{-1}\left[ x^1 j_2(\vx) - x^2 j_1(\vx)\right]=
\nonumber\\
&&=\frac{32\sin(F\pi)}{(4\pi)^{\frac{d+3}{2}}}\,
m^{\frac{d+1}{2}}\,r^{-\frac{d-1}{2}}
\int_1^\infty d v \; \sinh \left[ (2F-1)\Arch v \right]\;
v^{-\frac{d+1}{2}}\;K_{\frac{d+1}{2}}(2mr v).\qquad
\label{form29}
\end{eqnarray}
The calculation has been performed at $\Re\;d<3$, and then the result has been
continued analytically as a holomorphic function
on the whole complex $d$-plane. Unlike the
vacuum energy density (\ref{form24}), the vacuum current (\ref{form29})
vanishes at $F=\frac{1}{2}$, being negative and convex
downwards in the interval $0<F<\frac{1}{2}$, while positive and convex
upwards in the interval $\frac{1}{2}<F<1$; the positions of the minimum
and the maximum are symmetric with respect to the point $F=\frac{1}{2}$.
The asymptotics at large and small distances from the vortex are the
following:
\begin{equation}
j_{\varphi}(\vx)=\frac{2\sin(F\pi)}{(4\pi)^{\frac{d+1}{2}}}\left(F-\frac{1}{2} \right)
e^{-2mr} m^{\frac{d-3}{2}} r^{-\frac{d+3}{2}} 
\left\{ 1+O\left[(mr)^{-1}\right]\right\}, \, mr\gg 1,
\label{form30}
\end{equation}
\begin{eqnarray}
j_{\varphi}(\vx)=\frac{4\sin(F\pi)}{(4\pi)^{\frac{d}{2}+1}}
\left(F-\frac{1}{2} \right)
\frac{\Gamma\left(\frac{d-1}{2}+F\right)\Gamma\left(\frac{d-1}{2}+1-F\right)}
{\Gamma\left(\frac{d}{2}+1\right)}\, r^{-d} 
\left\{ 1+O\left[(mr)^2\right]\right\}, 
\nonumber\\mr\ll 1.\quad
\label{form31}
\end{eqnarray}

Note, that in the case of odd $d$ the integral in Eq.(\ref{form29})
can be taken, yielding a quadratic combination of the McDonald functions
of the variable $mr$, whereas in the case of even $d$ Eq.(\ref{form29})
can be transformed to a linear combination of the integral
$\displaystyle{\int_{2mr}^{\infty} \frac{du}{u} K_{2F-1}(u)}$ and the
McDonald functions of the variable $2mr$. In particular, we get
\begin{eqnarray}
&&j_{\varphi}(\vx)=\frac{\sin(F\pi)}{4\pi^2 r^2} \Big(F-\frac{1}{2}\Big)
\Bigg\{
-4\Big[ (F-\frac{1}{2})^2+m^2 r^2\Big]
\int_{2mr}^{\infty} \frac{du}{u} K_{2F-1}(u)+
\nonumber\\
&&+mr\Big[K_{2F}(2mr)+K_{2(1-F)}(2mr)\Big] \Bigg\}, \;\;\; d=2,
\label{form32}
\end{eqnarray}
\begin{eqnarray}
&&j_{\varphi}(\vx)=\frac{\sin(F\pi)}{6 \pi^3}\frac{m}{r^2}
\Bigg\{
\Big[ F(F-\frac{1}{2})+ m^2 r^2\Big]\,mr\,K_{F}^2(mr)-
\nonumber\\
&&
-\Big[ (1-F)(\frac{1}{2}-F)+m^2 r^2\Big]\,mr\,K_{1-F}^2(mr)+
\nonumber\\
&&
+2\Big[{F(1-F)}-m^2 r^2\Big]
(F-\frac{1}{2})K_{F}(mr)K_{1-F}(mr)
\Bigg\},\;\; d=3,
\label{form33}
\end{eqnarray}
\begin{eqnarray}
&&j_{\varphi}(\vx)=\frac{\sin(F\pi)}{32\pi^3 r^4} \Big(F-\frac{1}{2}\Big)
\Bigg(
4\bigg\{(F-\frac{1}{2})^2 \Big[ (F-\frac{1}{2})^2
-1+2m^2 r^2 \Big] + m^4 r^4 \bigg\}\times
\nonumber\\
&&\times
\int_{2mr}^{\infty} \frac{du}{u} K_{2F-1}(u)+2m^2 r^2 K_{2F-1}(2mr)-
\nonumber\\
&&-\Big[(F-\frac{1}{2})^2-1+m^2 r^2\Big]\,mr\,\Big[K_{2F}(2mr)+K_{2(1-F)}(2mr)\Big]
\Bigg), \;\; d=4,
\label{form34}
\end{eqnarray}
\begin{eqnarray}
&&j_{\varphi}(\vx)=\frac{\sin(F\pi)}{60 \pi^4}\frac{m}{r^4}
\Bigg(
\bigg\{ (F-\frac{1}{2})\Big[ (1+F)F(2-F)-(2F-\frac{1}{2})m^2 r^2\Big]-
\nonumber\\
&&- m^4 r^4 \bigg\}\,mr\,K_{F}^2(mr)
-\bigg\{ (\frac{1}{2}-F)\Big[ (1+F)(1-F)(2-F)-(\frac{3}{2}-2F)m^2 r^2\Big]-
\nonumber\\
&&- m^4 r^4 \bigg\}\,mr\,K_{1-F}^2(mr)
+2\bigg\{ (1+F)F(1-F)(2-F)+
\nonumber\\
&&+\Big[1-2F(1-F)\Big]m^2 r^2
+ m^4 r^4\bigg\}(F-\frac{1}{2})K_{F}(mr)K_{1-F}(mr) \Bigg),\;\; d=5;
\label{form35}
\end{eqnarray}
the result (\ref{form33}) has been already known for a rather long time
[9,10] \footnote{Note that
the results of Ref.\cite{Ser} concerning the vacuum energy density in the
case of $d=3$ are intrinsically controversial and completely wrong.}.

Supposing the validity of the Maxwell equation relating the vacuum current
to the vacuum magnetic field with the total flux (in the units of $2\pi$)
\begin{equation}
\Phi^{(I)}=\frac{e^2}{2}\int_{0}^{\infty} dr\, r^2 j_{\varphi}(\vx)
\label{form36}
\end{equation}
($e$ is the coupling constant possessing the dimension $m^{{\frac{3-d}{2}}}$),
we get
\begin{eqnarray}
\Phi^{(I)}=\frac{2\,e^2\, m^{d-3}}{3(4\pi)^{\frac{d+1}{2}}}
\Gamma\left(\frac{3-d}{2}\right) F\left(1-F\right)\left(F-\frac{1}{2}\right);
\label{form37}
\end{eqnarray}
this result is evaluated at $\Re\,d<3$ and can be continued analytically
as a meromorphic function on the whole complex $d$-plane. Note that in
the physically meaningful domain $d\ge2$ Eq.(\ref{form37}) , as well as
Eq.(\ref{form22}) at $s=-\frac{1}{2}$, has poles at odd $d$ and is
finite at even $d$.

Completing the analysis of the vacuum polarization effects in the background
of a singular magnetic vortex, we note that the vacuum energy density
(\ref{form24}) is even and the vacuum current (\ref{form29})  is
odd under charge conjugation. Note also that in the case of quantized massless
scalar field ($m=0$) the expressions for the vacuum energy density and current
are simplified considerably, see Eqs.(\ref{form21})  at $s=-\frac{1}{2}$
and (\ref{form31}).

In conclusion let us compare the above results with the pure Casimir effect,
i.e. the effect of the boundary condition (\ref{form4}) and (\ref{form5}) 
solely (in the absence of any background field) on the bosonic vacuum.
Certainly, the vacuum current is vanishing in this case, whereas
for the renormalized zeta function density we get the expression
\begin{equation}
\zeta_{\vx}^{ren}(s)=-\frac{1}{(2\pi)^{\frac{d}{2}}}
\frac{r^{2s-d}}{2^s \Gamma(s)} \int_{0}^{\infty} du\, u^{\frac{d}{2}-s-1}
\exp\left(-u-\frac{m^2 r^2}{2u}\right)I_{0}(u),
\label{form38}
\end{equation}
where $I_{\omega}(z)$ is the modified Bessel function of the order $\omega$.
The integral in Eq.(\ref{form38}) is divergent at $\Re\,s<\frac{d-1}{2}$
and cannot be extended analytically to this domain, thus the vacuum 
energy density (Eq.(\ref{form38})  at $s=-\frac{1}{2}$)
is infinite in the physically meaningful case $d \ge 2$.

\medskip
{\bf{Acknowledgements}}

The work was supported by the State Foundation for Fundamental Research
of Ukraine (project 2.4/320) and the Swiss National
Science Foundation (grant CEEC/ NIS/ 96-98/7 IP 051219).

\end {document}